\documentclass{article}
\usepackage{cite,spconf,amsmath,graphicx,enumerate,amssymb,color,bm}
\usepackage{algorithmic}
\usepackage{graphicx}
\usepackage{textcomp}
\usepackage{xcolor}
\usepackage{bbm}
\usepackage[caption=false, font=normalsize, labelfont=rm, textfont=rm]{subfig}

\def\x{{\boldsymbol x}}
\def\y{{\boldsymbol y}}
\def\z{{\boldsymbol z}}
\def\w{{\boldsymbol w}}

\def\s{{\boldsymbol s}}

\title{Versatile Semantic Coded Transmission over MIMO Fading Channels}

\name{Shengshi~Yao$^{\star}$,~Sixian~Wang$^{\star}$,~Jincheng~Dai$^{\star}$,~Kai~Niu$^{\star \dag}$,~Ping~Zhang$^{\star}$
\thanks{This work was supported in part by the National Natural Science Foundation of China under Grant 92067202, Grant 62001049, and Grant 62071058, in part by the Beijing Natural Science Foundation under Grant 4222012. \emph{(Corresponding author: Jincheng Dai)}}
}

\address{
\normalsize $^{\star}$ Beijing University of Posts and Telecommunications, Beijing, China\\
\normalsize $^{\dag}$ Peng Cheng Laboratory, Shenzhen, China\\
\normalsize Email: daijincheng@bupt.edu.cn\\
}

\begin{document}
\ninept
\maketitle
\begin{abstract}
Semantic communications have shown great potential to boost the end-to-end transmission performance. To further improve the system efficiency, in this paper, we propose a class of novel semantic coded transmission (SCT) schemes over multiple-input multiple-output (MIMO) fading channels. In particular, we propose a high-efficiency SCT system supporting concurrent transmission of multiple streams, which can maximize the multiplexing gain of end-to-end semantic communication system. By jointly considering the entropy distribution on the source semantic features and the wireless MIMO channel states, we design a spatial multiplexing mechanism to realize adaptive coding rate allocation and stream mapping. As a result, source content and channel environment will be seamlessly coupled, which maximizes the coding gain of SCT system. Moreover, our SCT system is versatile: a single model can support various transmission rates. The whole model is optimized under the constraint of transmission rate-distortion (RD) tradeoff. Experimental results verify that our scheme substantially increases the throughput of semantic communication system. It also outperforms traditional MIMO communication systems under realistic fading channels.
\end{abstract}


\section{Introduction}
Semantic communications are recently emerging as a new paradigm for data transmission \cite{zhang2022toward, niu2022paradigm, dai2022communication, xie2021deep, dai2022nonlinear}. Their superiority stems from semantics-guided joint source and channel design. Semantic coded transmission (SCT) lies at the heart of semantic communication systems, which are optimized end-to-end to achieve system performance gain \cite{bourtsoulatze2019deep, ding2021snr, kurka2021bandwidth, yang2022ofdm, yang2022deep, dai2022nonlinear}. Nevertheless, all existing works focus on SCT techniques using only one code stream transmitted over time and frequency domains. To further improve the transmission efficiency, for the first time, we introduce the multiple-input multiple-output (MIMO) techniques \cite{tse2005fundamentals} into SCT system to utilize an additional degree-of-freedom (DoF) provided by the spatial domain. In this way, the system throughput can be greatly improved. However, a naive combining of MIMO and SCT cannot meet expectations. Like that in traditional MIMO systems, one needs to develop proper channel coding and modulation strategy to each codeword stream for matching with the MIMO channel state. Herein, when MIMO is applied into the SCT system, one needs to elaborate a good match between MIMO channel and source semantics distribution.

To this end, we design a novel versatile SCT system over MIMO fading channels, named \emph{VST-MIMO}. The proposed framework features versatile rate transmission and multiple-stream transmission in parallel. Specifically, we design an adaptive spatial multiplexing (ASM) module to guide the rate allocation and stream mapping, coupling the source semantics and channel states. On one hand, a learnable entropy model is built to model the distribution of latent semantic features, indicating the source content. On the other hand, the rate allocation also depends on the channel condition, represented by channel quality indicator (CQI) for each stream. Accordingly, the proposed SCT framework achieves the dual adaptation to source semantics as well as channel states, and thus enables versatile transmission.

We verify the performance of the proposed VST-MIMO by simulations over image datasets. Compared to classical communication systems, as well as existing semantic communication systems, the proposed method achieves end-to-end rate-distortion (RD) performance gain. It substantially increases the throughput of the whole system and will catalyze the future application of semantic communications in reality.


\textit{Notational Conventions:} $p_x$ denotes a probability density function (pdf) with respect to the continuous-valued random variable $x$. $\mathbb{R}$ and $\mathbb{C}$ denote the real number set and the complex number set, respectively. $\mathcal{N}(x;\mu, \sigma^2) \triangleq (2\pi \sigma^2)^{-1/2} \exp(-(x - \mu)^2/(2\sigma^2))$ denotes a Gaussian function, and $\mathcal{CN}(x;\mu, \sigma^2)$ is the complex one. $\mathcal{U}(a-u,a+u)$ stands for a uniform distribution centered on $a$ with the range from $a-u$ to $a+u$. $\lfloor \cdot \rceil $ denotes scalar quantization, rounding to the nearest integer.

\section{System Model}\label{chap:method}

\subsection{Architecture}\label{sec:arch}

In this work, we consider a single-user MIMO communication scenario with $N_r$ receiving antennas and $N_t$ transmitting antennas at the user equipment (UE) and the base station (BS) respectively. The brief architecture of proposed semantic coded transmission framework for MIMO channels is displayed in Fig. 1, where $\left\{{\bm \phi}_g, {\bm \phi}_h, {\bm \phi}_f, {\bm \theta}_g, {\bm \theta}_h, {\bm \theta}_f \right\}$ encapsulates the learnable parameters of neural network functions.

\begin{figure}[t]
\label{Fig1}
\begin{center}
\centerline{\includegraphics[width=\columnwidth]{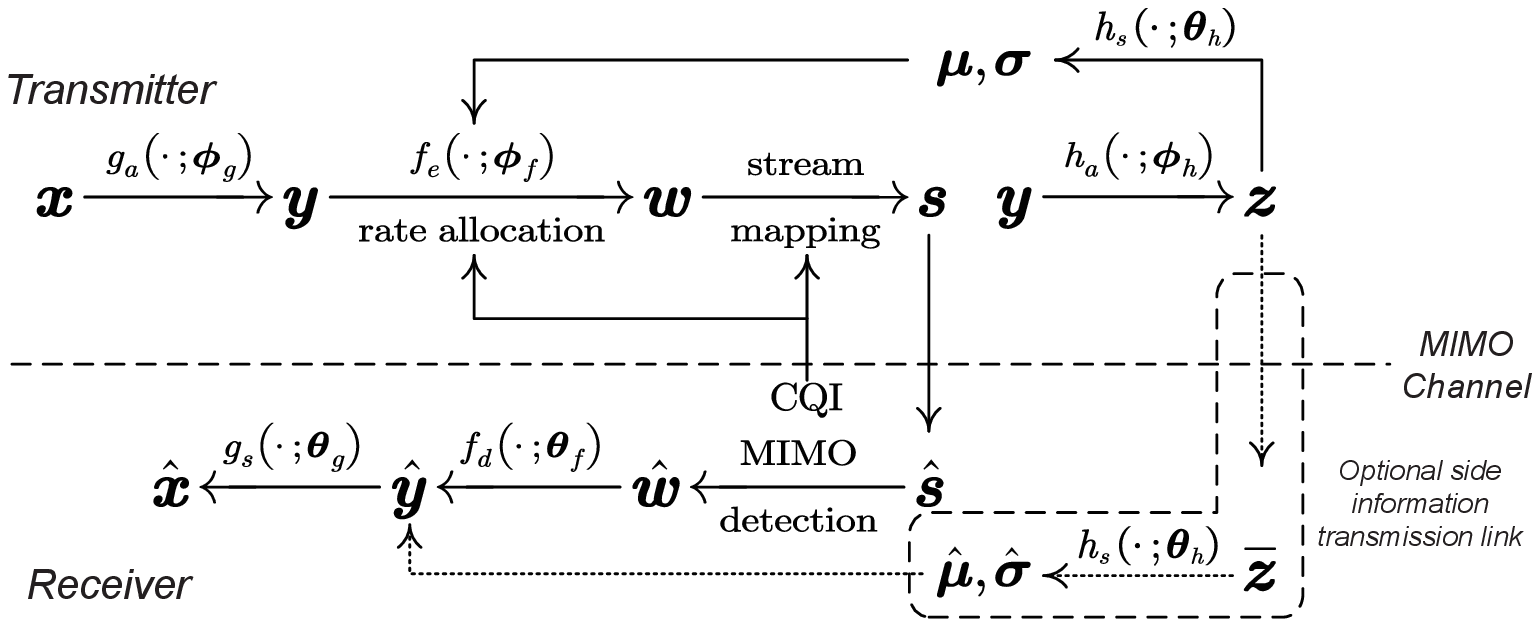}}
\vskip -0.15in
\caption{Overall architecture of proposed semantic coded transmission over MIMO fading channels. }
\vskip -0.4in
\label{Figsm}
\end{center}
\end{figure}

The analysis transform module $g_a$ transforms the input source vector $\x \in \mathbb{R}^m$ to its latent representation $\y$. From the latent code $\boldsymbol y$, it is firstly variationally modelled as a multivariate Gaussian. The hyperprior encoder $h_a$ summarizes the mean values $\boldsymbol{\mu}$ and standard derivations $\boldsymbol{\sigma}$ of $\boldsymbol y$ in the hyperprior $\z$. By means of entropy coding and channel coding, the quantized version $\bar \z =\lfloor \z \rceil $ is transmitted as side information over digital link. Secondly, the deep JSCC encoder $f_e$ encodes $\y$ as the symbol sequence $\boldsymbol w$. Specifically, an adaptive spatial multiplexing module (ASM) is designed to determine the length of each $w_i \in \boldsymbol w$, and map them into $N_s$ streams. Finally, $N_s$ symbol streams are distributed to $N_t$ physical antennas, yielding channel-input sequence $\s$. The detail of the ASM module is introduced in section \ref{sec:rate_adaptive}.

The channel input vector is transmitted through a MIMO channel. The received signal on the $c$-th subcarrier at the $j$-th receiving antenna is
\begin{equation}
  \hat{s}_{c,j} = \boldsymbol{h}_{c,j} \boldsymbol{p}_c  \boldsymbol{s}_c + n_{c,j},
\end{equation}
where $\boldsymbol{h}_{c,j} \in \mathbb{C}^{N_t}$ and $\boldsymbol{p}_c \in \mathbb{C}^{N_t\times \min(N_t,N_r)}$ denote the channel gain vector at the frequency domain and the precoding vector, respectively. $\boldsymbol{s}_c \in \mathbb{C}^{\min(N_t,N_r)\times 1}$ is the transmitted symbols on the $c$-th subcarrier, and $n_{c,j} \sim \mathcal{CN}(0,\sigma_n^2)$ is the additive Gaussian noise vector with noise power $\sigma_n^2$. Assuming the number of subcarriers is $N_c$, the CSI stacked in the frequency domain is formulated as $\boldsymbol{H} \in \mathbb{C}^{N_c \times N_r \times N_t}$, which is sampled from distribution $p_{\boldsymbol{H}}$. In this work, assuming CSI does not exist at the transmitter, we do not explicitly perform pre-coding, i.e., the pre-coded vector is jointly learned. The transmitter can acquire channel quality indicators (CQI), as input of the ASM module.

At the receiver side, following MIMO detection, deep JSCC decoder $f_d$ uses both the recovered symbols $\hat{ \boldsymbol{w}}$ and side information $\bar \z$ to estimate the latent representation $\y$ as $\hat{\y}$. Finally, the synthesis decoder $g_s$ reconstructs the source $\hat{\x}$ from $\hat{\y}$. Note that it is optional to transmit side information $\z$ in practice.

\subsection{Variational Modeling of Proposed Method}\label{sec:var_model}

Considering the spatial dependencies among the latent representation $\y$, following \cite{balle2018variational}, we introduce an additional set of latent variables $\z$ to represent the dependencies. In particular, each $y_i$ is variationally modeled as a Gaussian with mean ${\mu}_i$ and standard deviation ${\sigma}_i$, whose density function is factorized as
\begin{equation}\label{eq_ntscc_entropy_model}
  p_{\y|\z}(\y|\z) = \prod_i \mathcal{N}(y_i; {\mu}_i, {\sigma}_i^2) \text{~with~} (\bm{\mu},\bm{\sigma}) = {h_s}(\z; \bm{\theta}_h).
\end{equation}

To allow optimization via gradient descent in model training, as in \cite{balle2016end}, a proxy quantized representation $\tilde{\y} = \y + \bm{o} = g_{a,\bm{\phi}_g}(\x) + \bm{o}$ replaces $\bar \y=\lfloor \y \rceil $, where $\bm{o}$ is randomly sampled from standard uniform $\mathcal{U}(-\frac{1}{2}, \frac{1}{2})$. Hence, we derive a non-negative entropy estimation of $\y$ by convoluting $p_{\y|\z}$ with $\mathcal{U}(-\frac{1}{2}, \frac{1}{2})$, to guide the rate allocation in ASM module.
By using the deep JSCC encoder function $f_e$ and stream mapping,  $p_{\y|\z}(\y|\z)$ is further transformed to $p(\s|\z)$. Likewise, $\bar{\z}$ is replaced by the proxy quantization $\tilde{\z} = \z + \bm{o}$ during training. Since there is no prior information about $\tilde \z$, it can be modeled as fully factorized density \cite{balle2016end} as
\vspace{-0.5em}
\begin{equation}\label{eq_ntc_entropy_model_z}
  p_{\tilde \z| \bm{\psi}} (\tilde \z| \bm{\psi}) = \prod_j \left( p_{{z}_j | \bm{\psi}^{(j)}} ({z}_j | \bm{\psi}^{(j)}) * \mathcal{U}(-\frac{1}{2},\frac{1}{2}) \right) ({\tilde z}_j),
  \vspace{-0.6em}
\end{equation}
\noindent where $\bm{\psi}^{(j)}$ encapsulates all the parameters of $p_{{z}_j | \bm{\psi}^{(j)}}$ and ``$*$'' denotes the convolutional operation.

Combined with MIMO channel model, the variational inference computes
\vspace{-0.5em}
\begin{equation}\label{eq_ntscc_term1}
  q_{\hat \s,\tilde \z | \x} (\hat \s,\tilde \z | \x) = \prod_i p({\hat s}_i | \pi(\w), \boldsymbol{H}, \sigma_n) \cdot \prod_j \mathcal{U} ({\tilde z}_j | z_j - \frac{1}{2}, z_j + \frac{1}{2}),
\end{equation}
with $\y = g_a(\x; \bm{\phi}_g), \w = f_e(\y; \bm{\phi}_f), \z = h_a(\y; \bm{\phi}_h)$, and $\pi$ denotes stream mapping. The goal of the variance inference $q_{\hat{\s},\tilde{\z} | \x}$ is to approximate the intractable true posterior $p_{\hat{\s},\tilde{\z} | \x}$, by minimizing their KL divergence, i.e., $D_{\rm{KL}} \left[q_{\hat{\s},\tilde{\z} | \x} \| p_{\hat{\s},\tilde{\z} | \x} \right]$ over the data distribution $p_{\x}$ and CSI distribution $p_{\boldsymbol{H}}$. Finally, it turns out an RD optimization problem, i.e., minimizing

\vspace{-1em}
\begin{equation}\label{eq_ntscc_vae_target}
\begin{aligned}
\underset{\substack{\x \sim p_{\x}\\ \boldsymbol{H} \sim p_{\boldsymbol{H}}}}{\mathbb{E}} \underset{\hat \s, \tilde \z \sim q_{\hat \s, \tilde \z}}{\mathbb{E}} D_{\rm{KL}} \left[q_{\hat{\s},\tilde{\z} | \x} \| p_{\hat{\s},\tilde{\z} | \x} \right] \rightarrow \Big[\underbrace{- \log{p_{\tilde{\z}}(\tilde{\z})}}_{\text{rate of side info.}} \\
 \underbrace{- \log{p_{\hat{\s}|\tilde{\z}}(\hat{\s}|\tilde{\z})}}_{\text{transmission rate}} \underbrace{- \mathbb{E}_{\y\sim p_{\y | \hat{\s},{\tilde{\z}}}} \log p_{\x| {\y}}(\x| {\y}) \big]}_{\text{weighted distortion}} + \rm{const},
 \end{aligned}
 \end{equation}
where the first two terms quantify the total bandwidth cost and the third term represents the log-likelihood to recover $\x$. The detail of model training is introduced in section \ref{sec:training}.

\begin{figure*}[h]
\begin{center}
\centerline{\includegraphics[width=2\columnwidth]{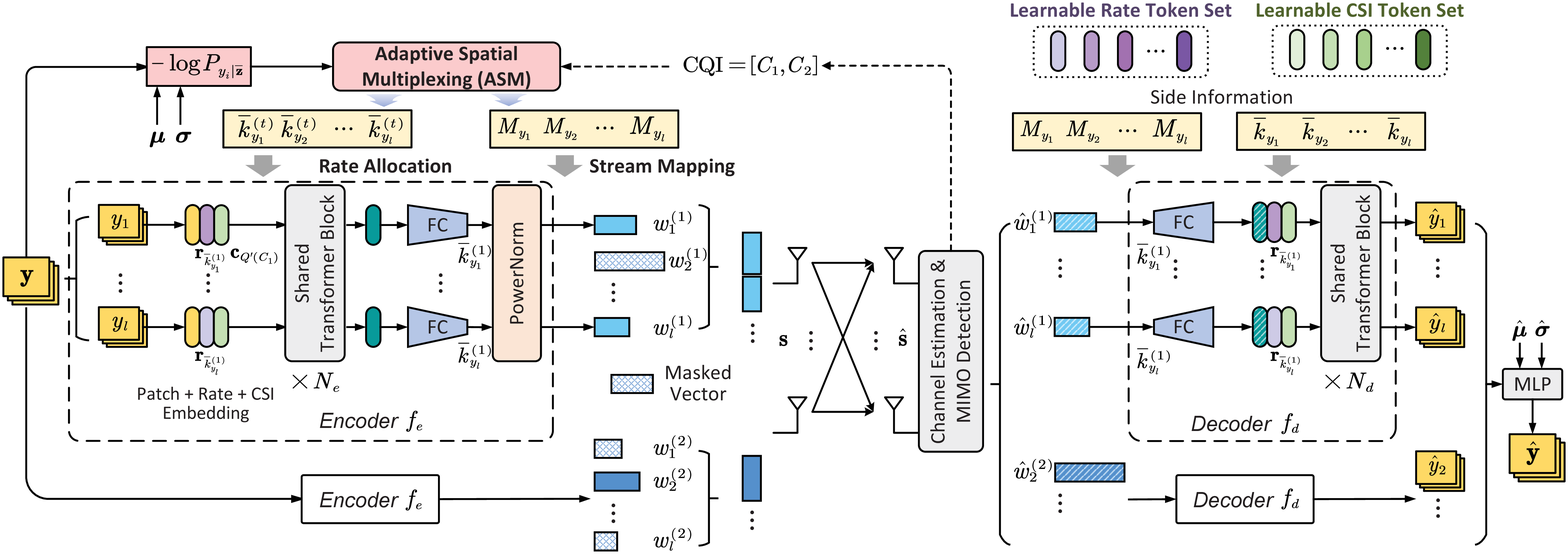}}
\caption{Versatile MIMO transmission architecture with adaptive rate allocation and stream mapping. FC denotes fully-connected network.}
\label{fig:versatile_transmission}
\end{center}
\vskip -0.2in
\end{figure*}

\subsection{Versatile Transmission over MIMO Fading Channels} \label{sec:rate_adaptive}

As indicated in section \ref{sec:arch}, we want to transmit the latent representation $\y$ of source data. It is fundamental to match the source contents and the channel environments, such as to obtain the end-to-end performance gain. Fig. \ref{fig:versatile_transmission} illustrates the versatile transmission mechanism with $N_s=2$ streams at large.
We design a unified pair of deep JSCC codecs, i.e., $f_e$ and $f_d$, to learn to transmit the latent representation $\y$ with variable transmission rates. The encoder $f_e$ firstly partitions the latent representation $\y$ into patch embedding sequence $\left(y_1, ..., y_i, . . . , y_l\right),i=1,2,. . .,l$, with $c$-dimensional vector each. Then, we propose an adaptive spatial multiplexing (ASM) module to guide the rate allocation and stream mapping.

The rate of transmitting $y_i$ is dual adaptive to two factors. On the one hand, the transmission rate depends on the entropy model
\begin{equation}
-\log P_{\tilde y_i|\boldsymbol{z}}(\tilde y_i|\z) = -\log (\mathcal{N}(y_i; {\mu}_i, {\sigma}_i^2) * \mathcal{U}(-\frac{1}{2},\frac{1}{2})),
 \end{equation}
which is the summation of entropy along all $c$ dimensions of $y_i$. 
On the other hand, the transmission rate is influenced by the CSI. In particular, the transmitter is informed of CQI for each stream, e.g., the signal-to-noise ratio (SNR) or signal-to-interference-plus-noise ratio (SINR) averaged over subcarriers in frequency domain, but the transmitter does not acquire the exact CSI vector. Accordingly, the equivalent averaged capacity $C_t$ is estimated for $t$-th stream. A larger $C_t$ allows for higher coding efficiency, i.e., less bandwidth cost. To sum up, if the latent feature vector $y_i$ is assigned to be transmitted in $t$-th stream, the cost for transmitting $y_i$, i.e., the length of $w_i$, can be formulated as

\begin{equation}
\label{eq:rate_control}
\bar{k}^{(t)}_{y_i} = Q\left(k^{(t)}_{y_i}\right)  = Q\left(-\frac{\eta}{C_t}\log p_{\tilde y_i|\z}\left(\tilde y_i|\z\right)\right),
\end{equation}
where $Q$ denotes a scalar quantizer with $2^{k_{q}} \text{ }(k_{q}=1,2,...)$ quantization levels and the quantization value set $\left\{\nu_1, \nu_2,...,\nu_{2^{k_q}}\right\}$ is predetermined. The averaged channel bandwidth cost can be flexibly controlled by adjusting $\eta$, which is a hyperparameter.

Moreover, the ASM module implements stream mapping according to source entropy and CSI. For convenience, let $M_{y_i} \in \left\{1,2,...,N_s\right\}$ denotes the index of the stream $y_i$ is assigned to, thus the total bandwidth cost $k_y$ to transmit $\y$ can be formulated as
\begin{equation}
\label{eq:total bandwidth}
k_y = \frac{N_s}{2N_t}\underset{t=1,2,...,N_s}{\max}\sum_{i=1}^l{\mathbbm{1} \left( M_{y_i}=t \right) \bar{k}_{y_i}^{(t)}},
\end{equation}
with two-dimensional constellations. $\mathbbm{1} \left( \cdot\right)$ is the indicator function. As we shall note, the cost of encoding $y_i$ is exactly $\bar{k}_{y_i}=\sum_{t=1}^{N_s} \mathbbm{1} \left( M_{y_i}=t \right) \bar{k}_{y_i}^{(t)}$.

As indicated in \cite{bourtsoulatze2019deep}, performance degrades when the CSI mismatches during training and inference. In order to make efficient utilization of deep JSCC codec and adapt to varying MIMO channel conditions, $f_e$ and $f_d$ are designed to adapt to various channel conditions. In particular, a rate token vector set $\mathcal{R} = \{ \mathbf{r}_{\nu_1}, \mathbf{r}_{\nu_2}, ..., \mathbf{r}_{\nu_{2^{k_q}}} \}$ is developed to indicate rate information, and
a CSI token vector set $\mathcal{C}=\left\{\mathbf c_{\tau_1}, \mathbf c_{\tau_2}, ..., \mathbf c_{\tau_I} \right\}$ to indicate CSI information. Another scalar quantizer $Q'$ with quantization values $\left\{\tau_1, \tau_2,...,\tau_I\right\}$ quantizes CQI value as $Q'(C_t)$. Combined with the rate token vector $\mathbf{r}_{\bar{k}^{(t)}_{y_i}}$ and CSI token vector $\mathbf{c}_{Q'(C_t)}$, $y_i$ is fed into Transformer blocks. Therefore, the Transformer blocks learn to adapt to the entropy of $y_i$ and the channel states, and then the following FC layer scales it to $\bar{k}^{(t)}_{y_i}$-dimensional vector $w^{(t)}_i$. In practice, in order to fully exploit the spatial dependencies among $\y$, all patch embeddings $\left\{y_i\right\}^{l}_{i=1}$ are fed into Transformer blocks, and $N_s$ streams can be encoded and decoded in parallel, yielding $\left\{w^{(t)}_i\right\}^{N_s}_{t=1}$. $w^{(t)}_i$ is retained for transmission only when $M_{y_i}=t$. Given the scaling factor $\eta$ and the total entropy of $\y$, $y_i$ with higher entropy is of higher priority and assigned to the stream with larger capacity, and so forth. Hence, for each $y_i$, additional $k_q+\log_2{N_s}$ bits are transmitted to inform the receiver which rate is allocated and from which stream to obtain $\hat{w}_i$.

\subsection{Modular Implementation Details}

We set up the analysis transform $g_a$ and the synthesis transform $g_s$ with the shifted-windows-based Swin Transformer \cite{liu2021swin}. An RGB image source $\x \in \mathbb{R}^{h\times w \times 3}$ is firstly divided into patches of $2 \times 2 \times 3$ dimensions. $g_a$ transforms these patches and outputs the latent representation $\y$, with $\y \in \mathbb{R}^{\frac{h}{4}\times \frac{w}{4} \times 3}$ for small-sized images and $\y \in \mathbb{R}^{\frac{h}{16}\times \frac{w}{16} \times 3}$ for large-sized ones. The synthesis transform $g_s$ has similar architecture designs with up-sampling instead.

\begin{figure*}[htbp]
    \vspace{-3em}
    \centering
    \hspace{-5em}
    \subfloat[]{
        \includegraphics[width=0.55\columnwidth]{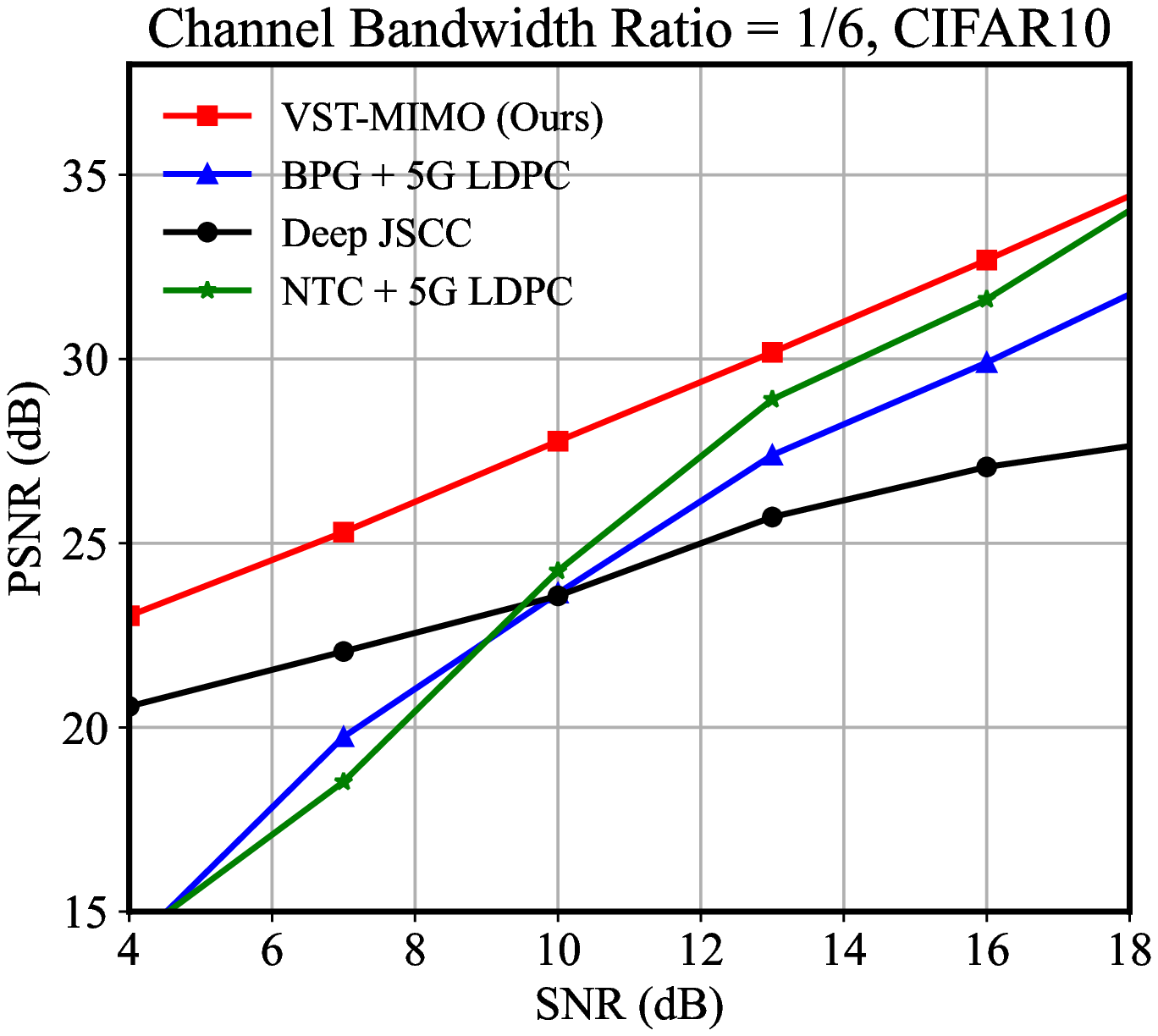}}
    \hspace{-1.7em}
    \subfloat[]{
        \includegraphics[width=0.55\columnwidth]{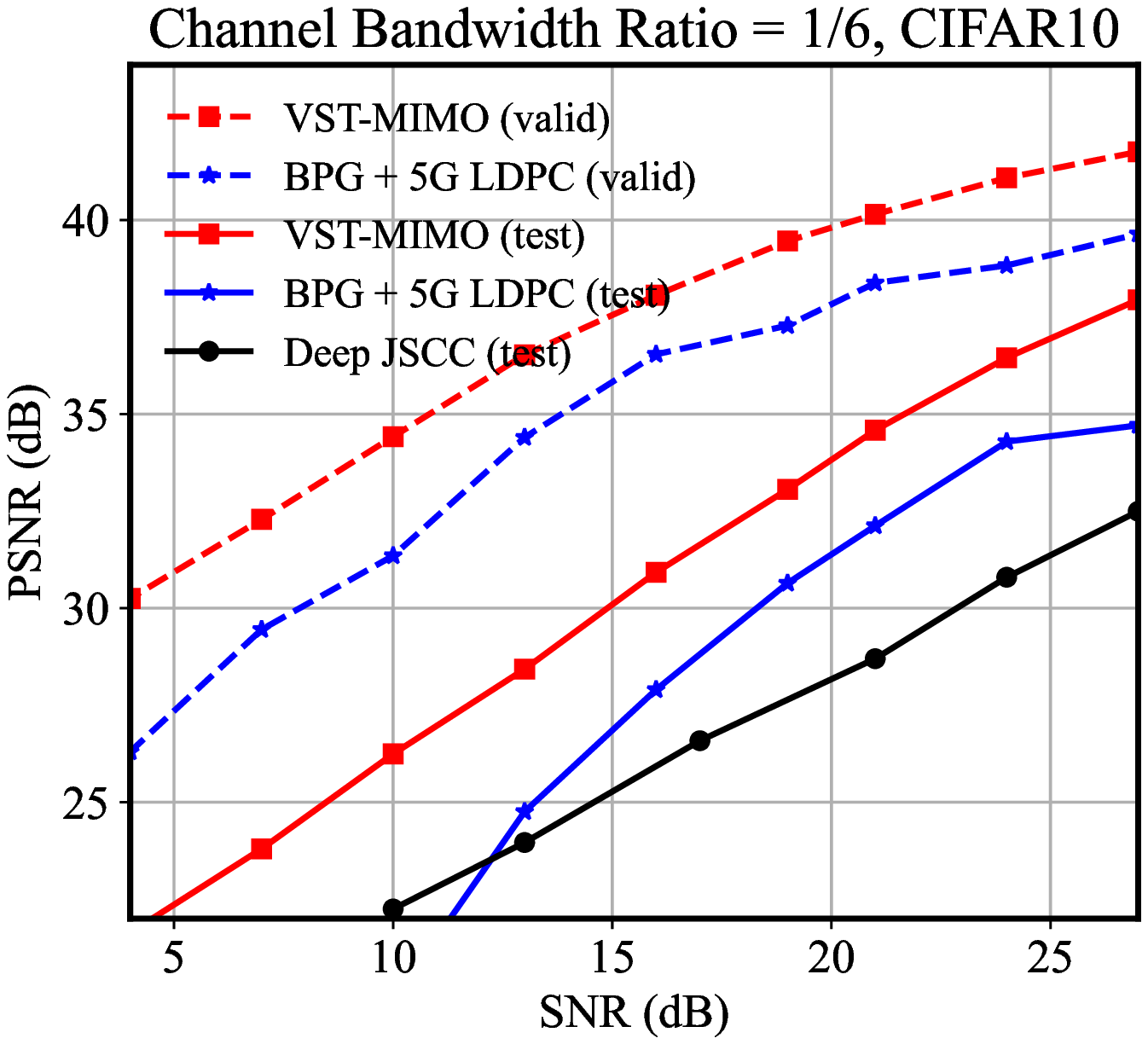}}
    \hspace{-1.7em}
    \subfloat[]{
        \includegraphics[width=0.55\columnwidth]{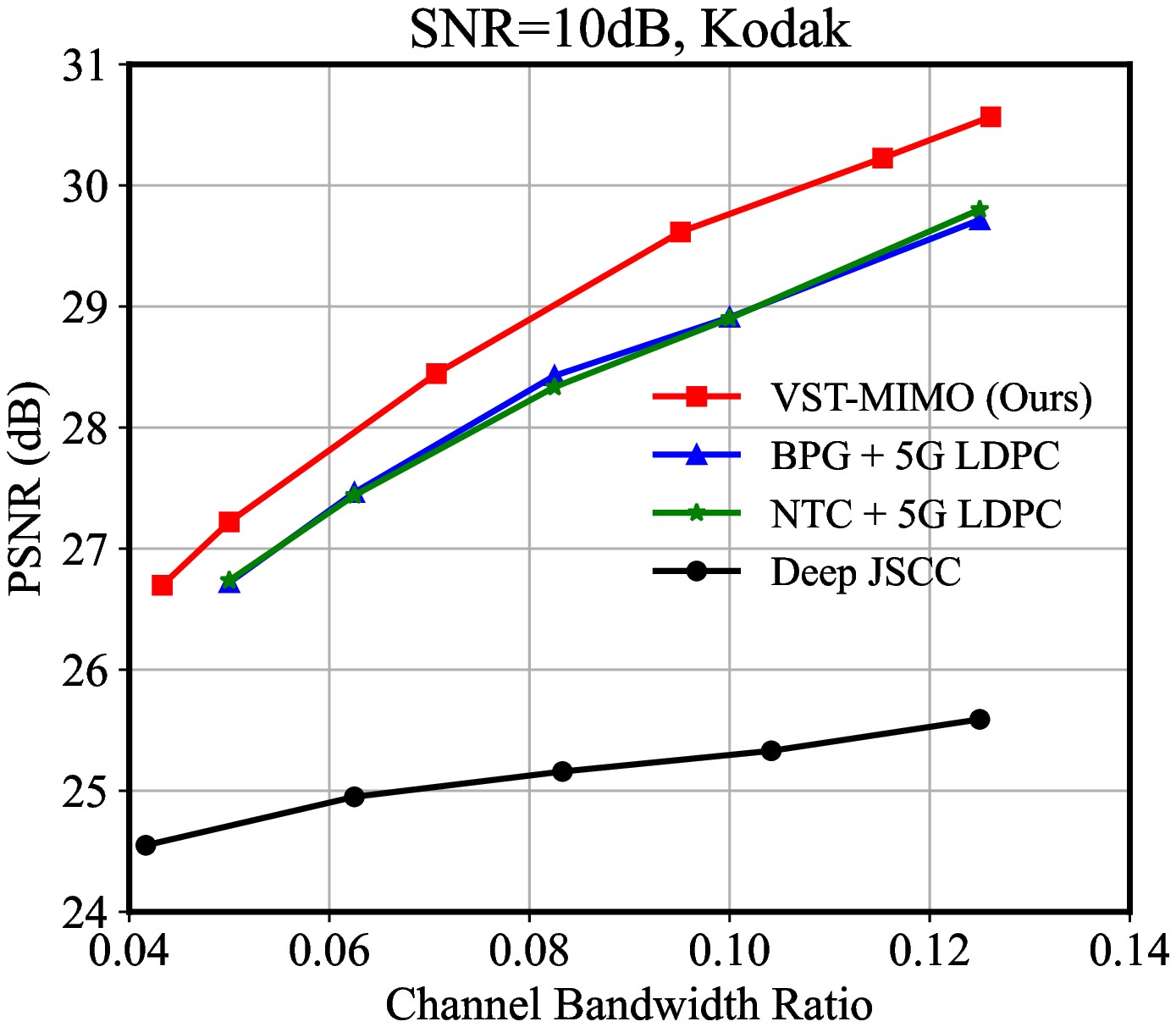}}
    \hspace{-1.5em}
    \subfloat[]{
        \includegraphics[width=0.55\columnwidth]{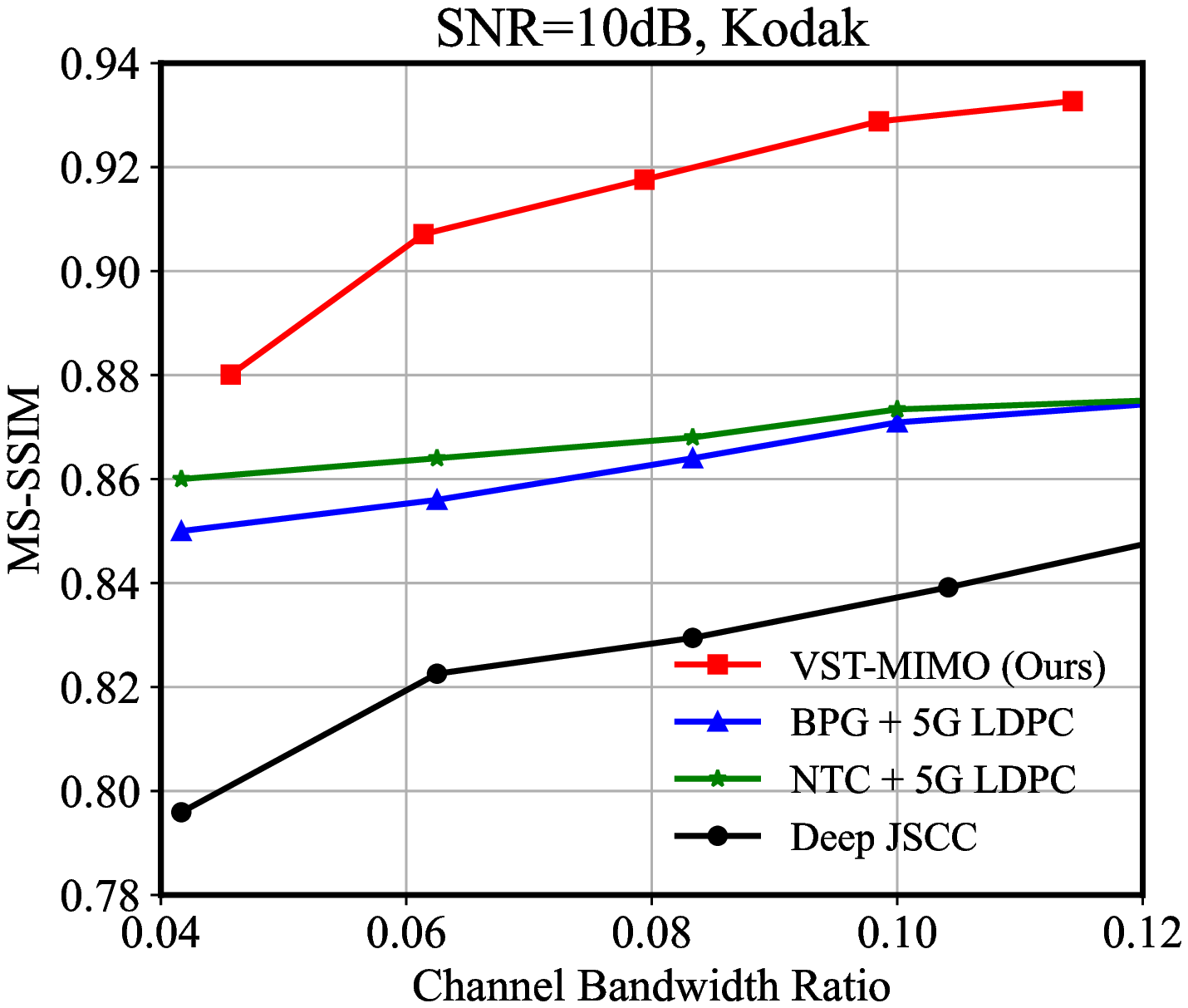}}
    \hspace{-5em}
    \label{fig3}
    \caption{(a)(b) PSNR performance versus SNR on CIFAR10 dataset over Kronecker channel and COST2100 channel.
    (c)(d) PSNR and MS-SSIM performance versus channel bandwidth ratio on Kodak dataset over COST2100 channel.}

\end{figure*}

At the receiver, streams are detected by classical zero-force MIMO detection algorithm, and then the reconstructed version $\hat{w}_i$ is obtained according to $M_{y_i}$.
We use $N_e = N_d = 4$ shared Transformer blocks in JSCC codec. Likewise, a bunch of FC layers decode $\{\hat{w}_i\}^{l}_{i=1}$ as vectors of unified dimensions. The aligned vectors are concatenated with the rate token and CSI token, and then fed to Transformer blocks. The hyperprior encoder $h_a$ and decoder $h_s$ are composed of three convolutional layers with ReLU activation function, with down-sampling and up-sampling each. If side information $\z$ is transmitted, $h_s$ also recovers $\hat{\boldsymbol{\mu}}$ and $\hat{\boldsymbol{\sigma}}$ from $\bar \z$, followed with a multi-layer perceptron (MLP) \cite{taud2018multilayer} to refine $\hat \y$.

\subsection{Model Training}\label{sec:training}

As discussed in section \ref{sec:var_model}, the minimizing of the KL divergence between the parametric variational density $q_{\hat{\s}, \tilde{\z} | \x}$ and the true posterior $p_{\hat{\s}, \tilde{\z} | \x}$ turns out an RD optimization problem. Substituting $p_{\tilde y_i|\z}$ with $p_{\tilde y_i|\tilde \z}$, the loss function $L$ can be formulated as a Lagrangian function, i.e.,
\vspace{-0.4em}
\begin{equation}\label{eq_ntscc_loss_function}
\begin{aligned}
  &L = \underset{\substack{\x \sim p_{\x}\\ \boldsymbol{H} \sim p_{\boldsymbol{H}}}}{\mathbb{E}} \bigg[ \lambda( \underbrace{-\sum_i {\frac{\eta}{C_{M_{y_i}}}\log p_{\tilde y_i|\boldsymbol{\tilde z}}\left(\tilde y_i|\boldsymbol{\tilde z}\right)}}_{\tilde k_y} \\
  &\underbrace{- \frac{\log{p_{\tilde \z| \bm{\psi}} (\tilde \z| \bm{\psi})}}{C_z}}_{\tilde k_z}) + d(\x,\hat \x)\bigg].
 \end{aligned}
  \vspace{-0.4em}
\end{equation}
$C_z$ denotes the digital channel capacity to transmit the quantized hyperprior $\bar \z$. Thus, the digital channel bandwidth cost $k_z$ can be computed to
transmit the side information. The Lagrange multiplier $\lambda$ controls the trade-off between the total bandwidth cost and the distortion $d$, which is mean square error function in our experiments.

  \vspace{-0.4em}
\section{Experiment}\label{chap:experiments}

  \vspace{-0.4em}

\subsection{Datasets and Experimental Setup}
We evaluate the end-to-end transmission performances over a small-size image dataset CIFAR10 \cite{Cifar10} ($32 \times 32$ pixels), and a medium-size dataset Kodak \cite{Kodak} ($768 \times 512$ pixels).

The number of antennas is set as $N_t=2$ at BS and $N_r=2$ at UE. The number of symbol streams is set as $N_s=2$. We consider two MIMO channel setups. Firstly, we consider a narrowband indoor MIMO channel named Kronecker model \cite{yu2001second}, whose channel covariance matrix can be well approximated by the Kronecker product of the covariance matrices from transmit and receive side, i.e., $\boldsymbol{R}_{\boldsymbol{H}}^T$ and $\boldsymbol{R}_{\boldsymbol{H}}^R$ respectively. Then, the channel gain matrix is generated by $\boldsymbol{H}=\left(\boldsymbol{R}_{\boldsymbol{H}}^R \right)^{1/2}\boldsymbol{G} \left(\boldsymbol{R}_{\boldsymbol{H}}^T\right)^{1/2}$, where elements of $\boldsymbol{G} $ are standard Gaussian complex random variables. In our experiments, we set $\boldsymbol{R}_{\boldsymbol{H}}^T = \left[ \begin{matrix}
	1&		0.2\\
	0.2&		1\\
\end{matrix} \right] $ and $\boldsymbol{R}_{\boldsymbol{H}}^R = \left[ \begin{matrix}
	1&		0.5\\
	0.5&		1\\
\end{matrix} \right] $.
Secondly, we consider the widely used COST2100 \cite{cost2100} as a wideband MIMO channel setup. The number of subcarriers is $N_c=1024$.

\subsection{Results}
The source vector $\x \in \mathbb{R}^m$ is transmitted with bandwidth cost $k=k_y+k_z$, and channel bandwidth ratio (CBR) \cite{DJSCCF} is defined as $R = k/m$. The rate-distortion performance is evaluated under peak signal-to-noise ratio (PSNR) and multi-scale structural similarity index measure (MS-SSIM, \cite{msssim}) metric.

To compare, we adopt BPG source coding \cite{BPG} combined with advanced low-density parity-check (LDPC) channel coding in 5G NR system \cite{3gpp} as the separate source and channel coding method, labeled as ``\emph{BPG + 5G LDPC}''. The LDPC code length is set as 4096. A proper modulation and coding scheme is selected according to the instantaneous quality of radio link, i.e., adaptive modulation and coding (AMC) \cite{3gpp}. Besides, we compare the scheme with nonlinear transform coding as source compression method, labeled as ``\emph{NTC + 5G LDPC}''. In this scheme, the latent representation $\y$ is quantized and entropy encoded according to the entropy model, followed by LDPC coding and modulation.
Apart from the comparison with separate coding schemes, we also compare our VST-MIMO with existing deep JSCC methods. The channel bandwidth is adjusted by using different numbers of convolutional kernels as in \cite{bourtsoulatze2019deep}.

Fig. 3(a) and Fig. 3(b) show the PSNR results of CIFAR10 dataset as a function of SNR (transmitter's SNR), with CBR constraint $R=1/6$, for Kronecker channel model and COST2100 model, respectively. Compared to standard deep JSCC, VST-MIMO bridges the gap between unbalanced source content and unbalanced channel states with the help of efficient rate allocation and stream mapping, thus improving the end-to-end performance significantly. It also shows competitive performance to the standard separate coding schemes with AMC. By directly encoding $\y$ without quantization, it provides additional performance gain compared to ``\emph{NTC + 5G LDPC}''. Fig. 3(b) also demonstrates the generalization ability under a test channel environment featured with $p'_{\boldsymbol{H}}$ (solid lines, test), while the model is only trained over $p_{\boldsymbol{H}}$ (dashed lines, valid).

We further plot the PSNR and MS-SSIM results versus the channel bandwidth ratio at $\text{SNR}=10\text{dB}$ over COST2100 MIMO channel in Fig. 3(c) and Fig. 3(d). In particular, our proposed VST-MIMO can save up to $24\%$ bandwidth cost compared to separative coding schemes when achieving the same PSNR. Besides, the proposed model outperforms the competitors in objective perceptual metric MS-SSIM (1.0 is best) by a large margin at various CBRs. In other words, the system throughput is increased with the same transmission performance.
As we shall note, with the help of the dual adaptive rate transmission mechanism, the model is empowered with versatile transmission integrated into one model, while one standard deep JSCC model is specific for single rate option.

  \vspace{-0.4em}
\section{Conclusion}\label{chap:conclusions}

  \vspace{-0.4em}
This paper presents a versatile semantic coded transmission architecture VST-MIMO over wireless MIMO fading channels. The adaptive spatial multiplexing mechanism enables the system to be dual adaptive to the source content and channel states, thus supporting multi-stream cooperative transmission. The proposed model achieves substantial gain under the established metrics and shows great potential in future semantic communications.

\vfill
\pagebreak

\clearpage

\bibliographystyle{IEEEbib}
\bibliography{bliography}

\end{document}